\documentclass[prl,twocolumn]{revtex4}
\usepackage[T1]{fontenc}
\usepackage[latin9]{inputenc}
\usepackage{verbatim}
\usepackage{float}
\usepackage{amsmath}
\usepackage{graphicx}
\usepackage{esint}

\makeatletter

\providecommand{\tabularnewline}{\\}

\@ifundefined{textcolor}{}
{%
 \definecolor{BLACK}{gray}{0}
 \definecolor{WHITE}{gray}{1}
 \definecolor{RED}{rgb}{1,0,0}
 \definecolor{GREEN}{rgb}{0,1,0}
 \definecolor{BLUE}{rgb}{0,0,1}
 \definecolor{CYAN}{cmyk}{1,0,0,0}
 \definecolor{MAGENTA}{cmyk}{0,1,0,0}
 \definecolor{YELLOW}{cmyk}{0,0,1,0}
}

\makeatother

\begin{document}

\title{Generalization of the Wall Theorem to Out-of-equilibrium Conditions}

\author{Ignacio Urrutia$^{1,2,}$}
\email{iurrutia@cnea.gov.ar}

\author{Iván E. Paganini$^{1,2}$}

\author{Claudio Pastorino$^{1,2}$}

\affiliation{$^{1}$Departamento de Física de la Materia Condensada, Centro Atómico
Constituyentes,CNEA, Av.Gral.~Paz 1499, 1650 Pcia.~de Buenos Aires,
Argentina}

\affiliation{$^{2}$Instituto de Nanociencia y Nanotecnología, CONICET-CNEA, CAC.}
\begin{abstract}
The well-known Wall theorem states a simple and precise relation among
temperature, pressure and density of a fluid at contact with a confining
hard wall in thermodynamic equilibrium. In this Letter we develop
an extension of the Wall theorem to out-of-equilibrium conditions,
providing an exact relation between pressure, density and temperaure
at the wall, valid for strong non-equilibrium situations. We derive
analytically this Non-equilibrium Wall theorem for stationary states
and validate it with non-equilibrium event-driven molecular-dynamics
simulations. We compare the analytical expression with simulations
by direct evaluation of temperature, density and pressure on the wall
in linear regime, medium and very strong out-of-equilibrium conditions
of a nanoconfined liquid under flow in stationary state, presenting
viscous heating and heat transport. The agreement between theory and
simulation is excellent, allowing for a conclusive validation. In
addition, we explore the degree of accuracy of using the equilibrium
Wall theorem and different expressions for the local temperature,
employed in non-equilibrium molecular-dynamics simulations.
\end{abstract}
\maketitle

Very few exact relationships are available in the context of inhomogeneous
fluids. One of such is the contact theorem also known as the Wall
theorem (WT), which establishes an ideal-gas-like equation of state
for the intensive properties of a fluid in thermodynamic equilibrium\emph{,}
at contact with a wall. The WT applies to fluids confined by hard
walls that constrain the region of the space occupied by the fluid
through an external zero-infinite potential. The first version of
the WT\citep{Lebowitz_1960} was derived for a fluid in contact with
a planar wall (i.e. filling a half-space) and reads
\begin{equation}
P_{w}=\rho_{w}kT\:,\label{eq:wt}
\end{equation}
where $P_{w}$ is the pressure on the wall, $\rho_{w}$ is the density
at contact with the wall, $T$ is the temperature of the system and
$k$ the Boltzmann constant. In this particular case, the geometry
of the system imposes also the extra condition $P_{w}=P$, being $P$
the pressure of the bulk fluid.\citep{Lebowitz_1960,Henderson_1984,HendersonD_1987}

The WT, expressed in Eq. (\ref{eq:wt}), also concerns to fluids constrained
by curved walls as spheres and cylinders.\citep{Henderson_1983,Henderson_1986_inCroxton1986,Henderson_1984_b}
For the application of Eq. (\ref{eq:wt}) to non-planar walls, it
becomes necessary to introduce the dividing surface, which fixes the
position of the surface where the pressure of the fluid acts on the
vessel walls. As a matter of fact, Eq. (\ref{eq:wt}) is the expression
of the WT for constant-curvature surfaces, written for a dividing
surface which coincides with the position of the discontinuity of
the hard wall-fluid potential. The choice for the location of the
dividing surface can be different, but a transformation between them
is straightforward.\citep{Urrutia_2014,Reindl_2015} WT applies, for
example, to fluids confined in pores both of simple\citep{Lutsko_2010,Kim_2012_b,Paganini_2015}
and complex,\citep{Urrutia_2010b} shapes. It was used under a broad
variety of conditions, spaning from large systems, in the thermodynamic
limit, to fluid-like very small systems comprised of very few particles.\citep{Urrutia_2010b,Urrutia_2011_b}
Key aspects of the curvature dependence of the surface free energy
in confined fluids were revealed using the WT by adopting different
approaches as density functional theories,\citep{Blokhuis_2007,Blokhuis_2013,HansenGoos_2014}
molecular dynamics\citep{Paganini_2015} and virial series.\citep{Urrutia_2014,Urrutia_2016b}
Several formulations of the WT have been developed along time. For
example, a WT was postulated for fluids composed by charged particles
in contact with charged walls.\citep{HendersonD_1979,Bhuiyan_2008,HendersonD_2009}

The interest in the behavior of nanoconfined fluids is evident in
the fields of Micro- and Nanofluidics and in technological applications.
Efficient use of fluids for heat removal or good thermal isolation
at small scales, is crucial to advance in the miniaturization of current
technical developments.\citep{Schoch_08,Squires_05} The interrelation
between flow regime and adsorption of gas confined in nanoporous shale
is actively studied to optimize natural gas production from shale
gas reservoirs.\citep{Kazemi_2016,Wu_2017b,Shen_2018} Highly confined
inhomogeneous liquids with a relatively high surface-to-volume ratio
and/or under flow, are common physical situations in those areas and
the interface of the fluid with the confining wall is a key aspect
to understand and tailor.\citep{Hu_17} In this work, we are interested
in hard walls that induce their temperature to the confined fluid.
Naturally, the thermal hard wall cannot describe in detail the properties
of a real substrate. Even though, the relevance of the thermal hard
wall model to study the fluxes of energy and mass in confined fluids
relies in that it is a simple prescription allowing to study the system
under minimal assumptions and direct analytical calculations.

In this letter we demonstrate the Non-equilibrium Wall theorem (NEWT),
applied to stationary states with thermal and velocity gradients.
Under flow, viscous heating and heat transfer scenarios, we also test
the NEWT using event-driven molecular-dynamics simulations (MD) of
fluid flow through a narrow cylindrical channel. Our results show
the validity of NEWT from near-equilibrium up to very strong non-equilibrium
conditions.

Consider a fluid system with fixed number of particles $N$ each with
mass $m$, in a confined region $\mathcal{A}$ with volume $V$. $\mathcal{A}$
is enclosed by a wall or substrate which is at a temperature $T_{{\rm wall}}$.%
{} If such a fluid system is in equilibrium, its temperature is everywhere
$T=T_{{\rm wall}}$ and the partition function for the system reads
\[
Q=\iint f\left(\mathbf{r}^{N},\mathbf{p}^{N}\right)d\mathbf{r}^{N}d\mathbf{p}^{N}\:,
\]
where $f\left(\mathbf{r}^{N},\mathbf{p}^{N}\right)=C\exp[-(\phi+\psi+K)/kT]$,
with $K$ the kinetic energy, $\phi$ the external potential and $\psi$
the interaction potential between particles. This function can be
factorized in different exponential terms. We focus on the external
potential term $\exp[-\phi/kT]$ for the case of a hard-wall external
potential $\phi$. Now, we introduce the boundary indicator function
$f^{(b)}\equiv\exp[-\phi/kT]=\prod_{i}^{N}\varTheta\left(-\left|\mathcal{A}-\mathbf{r}_{i}\right|\right)$
with $\varTheta$ the Heaviside function and $\left|\mathcal{A}-\mathbf{r}\right|$
the (shortest) distance between point $\mathbf{r}$ and $\mathcal{A}$,
which is zero only if $\mathbf{r}\in\mathcal{A}.$ Here, $\varTheta\left(x\right)=1$
if $x\geq0$ and $\varTheta\left(x\right)=0$ if $x<0$. Naturally,
we define $V=\intop\varTheta\left(-\left|\mathcal{A}-\mathbf{r}\right|\right)d\mathbf{r}$
and $A=\intop\delta\left(\left|\partial\mathcal{A}-\mathbf{r}\right|\right)d\mathbf{r}$,
with $A$ the surface area of $\partial\mathcal{A}$, the boundary
of the system and $\delta$ stands for the Dirac delta function. The
wall not only determines the boundary of the system, but it is also
a thermal wall, which sets the temperature of the bouncing particles.

For non-equilibrium (NE) stationary conditions we write 
\[
Q=\iint f_{\mathrm{NE}}\left(\mathbf{r}^{N},\mathbf{p}^{N}\right)d\mathbf{r}^{N}d\mathbf{p}^{N}\:,
\]
where $f_{{\rm NE}}$ can be factorized in different terms. One of
them includes the hard wall external potential and is still given
by $f^{(b)}$, which fixes the boundary of the system. Now, the fluid
could have different temperatures, in different regions, but the thermal
wall acts as a thermal reservoir which induces locally its temperature
to the fluid. Essentially, the thermal wall affects the temperature
of particles once they bounce on the wall. As regards non equilibrium
features, the statistical distributions are in general, of course,
non-trivial. We divide the complete domain $\left(\mathbf{r},\mathbf{p}\right)$
in different subsets. For our purposes it is convenient to split the
momentum space among opposite directions $\hat{\mathbf{n}}_{{\rm out}}=\hat{\mathbf{n}}$
and $\hat{\mathbf{n}}_{{\rm in}}=-\hat{\mathbf{n}}$, where \textbf{$\hat{\mathbf{n}}$}
is the normal versor to $\partial\mathcal{A}$ pointing to the outward
direction of $\mathcal{A}$. $\hat{\mathbf{n}}$ depends on $\mathbf{r}\in\mathcal{A}$.
Thus, we introduce the position-momentum one body density distributions
\begin{eqnarray}
\rho_{s}\left(\mathbf{r},\mathbf{p}\right) & = & Q^{-1}\iint\sum_{i}^{N}\varTheta\left(\mathbf{p}_{i}\cdot\hat{\mathbf{n}}_{s}\right)\delta\left(\mathbf{p}-\mathbf{p}_{i}\right)\times\nonumber \\
 &  & \delta\left(\mathbf{r}-\mathbf{r}_{i}\right)f_{\mathrm{NE}}\left(\mathbf{r}^{N},\mathbf{p}^{N}\right)d\mathbf{r}^{N}d\mathbf{p}^{N}\:,\label{eq:rhorp}
\end{eqnarray}
and $\rho\left(\mathbf{r},\mathbf{p}\right)=\rho_{{\rm in}}\left(\mathbf{r},\mathbf{p}\right)+\rho_{{\rm out}}\left(\mathbf{r},\mathbf{p}\right)$.
The position-dependent number density distributions are $\rho_{s}\left(\mathbf{r}\right)=\int\rho_{s}\left(\mathbf{r},\mathbf{p}\right)d\mathbf{p}$
with $N_{s}=\int\rho_{s}\left(\mathbf{r}\right)d\mathbf{r}$, and
subsystem index $s={\rm in},{\rm out}$ with in and out velocities,
respectively. Furthermore, the usual number density is $\rho\left(\mathbf{r}\right)=\sum_{s}\rho_{s}\left(\mathbf{r}\right)=\rho_{{\rm in}}\left(\mathbf{r}\right)+\rho_{{\rm out}}\left(\mathbf{r}\right)$.
Variables without subscript correspond to the total system. Eq. (\ref{eq:rhorp})
is the equivalent to the statistical-mechanical definition $\rho_{s}\left(\mathbf{r},\mathbf{p}\right)=\left\langle \sum_{i}^{N}\delta\left(\mathbf{r}-\mathbf{r}_{i}\right)\delta\left(\mathbf{p}-\mathbf{p}_{i}\right)\right\rangle _{s}$.
We note that in NE conditions it could be the case that $\rho_{{\rm in}}\left(\mathbf{r}\right)\neq\rho_{{\rm out}}\left(\mathbf{r}\right)$,
but at equilibrium, the relation $\rho_{{\rm in}}\left(\mathbf{r}\right)=\rho_{{\rm out}}\left(\mathbf{r}\right)=\rho\left(\mathbf{r}\right)/2$
must hold.

Number densities and other mean magnitudes of the fluid on the wall,
labeled as $w$, can be written as:
\begin{eqnarray}
\rho_{w,s} & = & A^{-1}\!\!\iint\rho_{s}\left(\mathbf{r},\mathbf{p}\right)\delta\left(\left|\partial\mathcal{A}-\mathbf{r}\right|\right)d\mathbf{r}d\mathbf{p}\:,\label{eq:rhows}\\
\left[X\right]_{s}\rho_{w,s} & = & A^{-1}\!\!\iint X\rho_{s}\left(\mathbf{r},\mathbf{p}\right)\delta\left(\left|\partial\mathcal{A}-\mathbf{r}\right|\right)d\mathbf{r}d\mathbf{p}\:,\label{eq:Xws}
\end{eqnarray}
where $X$ is a function of one particle position and momentum $\left(\mathbf{r},\mathbf{p}\right)$.
For example, the mean velocity in a tangential direction $\hat{\mathbf{t}}$
of particles coming out of the wall is $\left[\mathbf{p}\cdot\hat{\mathbf{t}}\right]_{{\rm out}}/m$.
The total mean value on the wall is recovered by the expression $\left[X\right]=\left(\left[X\right]_{{\rm in}}\rho_{w,{\rm in}}+\left[X\right]_{{\rm out}}\rho_{w,{\rm out}}\right)/\rho_{w}$.

We point out that in non-equilibrium conditions different forms of
measuring the temperature may give different results.\citep{Popov_2007,Patra_2017}
A frequently used prescription for calculating the temperature in
MD simulations is through the square of the particles velocity, relative
to the stream velocity. It is referred as the \emph{kinetic} temperature.\citep{Liu_2010,frenkel-smit}
On the other hand, we can evaluate the temperature using the kinetic
energy in different characteristic directions that do not present
a net flux of particles.\citep{Todd2017} Here, we follow this approach
to introduce the temperature measured in the normal direction
\begin{equation}
kT_{w,s}^{(n)}m=\left[\left(\mathbf{p}\cdot\hat{\mathbf{n}}\right)^{2}\right]_{s}\:,\label{eq:kTnws}
\end{equation}
associated with the velocity normal to the wall.%
{} We may note that, as it could be the case that $\rho_{w,{\rm in}}\neq\rho_{w,{\rm out}}$,
then it could also happen that the temperature of the fluid on the
wall, $T_{w,{\rm in}}^{(n)}$ and $T_{w,{\rm out}}^{(n)}$, were different
to $T_{{\rm wall}}$. Alongside, if equilibrium is established there
is a unique temperature for the entire system $T=T_{{\rm wall}}=T_{w,{\rm in}}^{(n)}=T_{w,{\rm out}}^{(n)}$.

The pressure on the wall is the force exerted by the fluid per unit
area and normal to the wall. It results from the mean value of momentum
transfer (flux) between the fluid and the wall substrate in the normal
direction. Taking into account the number of incident (${\rm in}$)
and scattered (${\rm out}$) particles with normal velocity $\mathbf{v}\cdot\hat{\mathbf{n}}$
and the momentum they transfer to the wall, we obtain
\begin{equation}
P_{w,s}=A^{-1}\!\!\iint\frac{\left(\mathbf{p}\cdot\hat{\mathbf{n}}\right)^{2}}{m}\rho_{s}\left(\mathbf{r},\mathbf{p}\right)\delta\left(\left|\partial\mathcal{A}-\mathbf{r}\right|\right)d\mathbf{r}d\mathbf{p}\:.\label{eq:Pws}
\end{equation}
Expression (\ref{eq:Pws}) requires a derivation that will be given
in the next paragraph. A comparison between Eqs. (\ref{eq:Xws},
\ref{eq:kTnws}) and (\ref{eq:Pws}) shows that normal kinetic temperature
$T_{w,s}^{(n)}$ plays a special role. Here we collect results from
Eq. (\ref{eq:rhows}) to (\ref{eq:Pws}), to obtain that 
\begin{equation}
P_{w,s}=\rho_{w,s}kT_{w,s}^{(n)}\:.\label{eq:NEWTs}
\end{equation}
It addition, it can be verified that the total momentum flux corresponds
to the sum over $s$-index, taking the values ${\rm in}$ and ${\rm out}$.
Then, the pressure on the wall is $P_{w}=P_{w,{\rm out}}+P_{w,{\rm in}}$.
In this way we obtain the mean result of the present work:
\begin{equation}
P_{w}=\rho_{w}kT_{w}^{(n)}\:.\label{eq:NEWT}
\end{equation}
Eq. (\ref{eq:gNEWT}) expresses the generalization to non-equilibrium
conditions of the Wall theorem for the pressure. It simply states
that under NE, even when the different forms of measuring the kinetic
temperature are non-equivalent, the WT still applies, if the temperature
$T$ is replaced by the kinetic temperature measured in the direction
normal to the wall.

\textbf{Derivation:} Instead of focusing on the momentum flux given
in Eq. (\ref{eq:Pws}), we analyze now a more general case: the flux
of a generic quantity $Y$. We consider the skin $\partial\mathcal{A}^{\epsilon}$
of $\mathcal{A}$, that extends a small depth $\epsilon$ from $\partial\mathcal{A}$,
towards the inner direction, and a small interval of momentum in the
normal direction $(p_{z},p_{z}+\Delta p_{z})$. The normal outward
versor for any point in $\partial\mathcal{A}^{\epsilon}$ is $\hat{z}\equiv\hat{\mathbf{n}}_{{\rm out}}$.
Each particle in $\partial\mathcal{A}^{\epsilon}$ has a constant
velocity to a good approximation. This approximation becomes better
for a progressively smaller $\epsilon$. The number of particles
in $\partial\mathcal{A}^{\epsilon}$ that will collide with the wall
per unit time, having normal momentum between $p_{z}$ and $p_{z}+\Delta p_{z}$
along an small time interval $\tau$ is given by
\[
\frac{1}{\tau}\iint\!\int_{-\tau p_{z}/m}^{0}\!\!\rho_{{\rm out}}\left(\mathbf{r},\mathbf{p}\right)J_{\mathbf{r}}J_{\mathbf{p}}dz\Delta p_{z}d\mathbf{r}_{\bar{z}}d\mathbf{p}_{\bar{z}}\,.
\]
We write $d\mathbf{r}=J_{\mathbf{r}}dzd\mathbf{r}_{\bar{z}}$ to explicitly
separate the differential in $z$ direction from the other spatial
directions packed in $d\mathbf{r}_{\bar{z}}$, with $J$ the jacobian,
and assume $\epsilon>\tau p_{z}/m$. The value $z=0$ is the wall
position. The mean flux of $Y\left(\mathbf{r},\mathbf{p}\right)$
is then
\begin{equation}
\frac{1}{\tau}\iint\!\int_{-\tau p_{z}/m}^{0}\!\!Y\rho_{{\rm out}}\left(\mathbf{r},\mathbf{p}\right)J_{\mathbf{r}}J_{\mathbf{p}}dz\Delta p_{z}d\mathbf{r}_{\bar{z}}d\mathbf{p}_{\bar{z}}\,.\label{eq:ded01}
\end{equation}
Specifically, this is the flux of $Y$ towards the wall exerted by
particles with normal momentum in the range $\left(p_{z},p_{z}+\Delta p_{z}\right)$.
Now we integrate in $dz$, after that we take the limits $\underset{\epsilon\rightarrow0}{\lim}\,\underset{\tau\rightarrow0}{\lim}$
and finally we replace $\Delta p_{z}$ by $dp_{z}$ to integrate in
$dp_{z}$, which reduces Eq. (\ref{eq:ded01}) to 
\begin{equation}
\iint\int\frac{\mathbf{p}\cdot\hat{\mathbf{n}}}{m}Y_{w}\left(\mathbf{p}\right)\rho_{w,{\rm out}}\left(\mathbf{p}\right)\left(J_{\mathbf{r}}J_{\mathbf{p}}\right)_{w}dp_{n}d\mathbf{r}_{\bar{n}}d\mathbf{p}_{\bar{n}}\,.\label{eq:ded02}
\end{equation}
We cast this to a volumetric integral in space, following Eq. (\ref{eq:Xws}),
and collect the result for both directions, $s={\rm in}$ and $s={\rm out}$,
to obtain
\begin{equation}
j_{w,s}\!\left(Y\right)=\frac{1}{A}\!\!\iint\!\frac{\mathbf{p}\cdot\hat{\mathbf{n}}}{m}Y\rho_{s}\left(\mathbf{r},\mathbf{p}\right)\delta\left(\left|\partial\mathcal{A}-\mathbf{r}\right|\right)d\mathbf{r}d\mathbf{p}\,.\label{eq:jY}
\end{equation}
$j_{w,s}\!\left(Y\right)$ is the flux of $Y$ per unit time and area
between the fluid and the wall for the $s$-subsystem, in the direction
normal to the wall. The deduction for $s={\rm in}$ is presented in
the Supplementary Material (SM). Collecting results from Eqs. (\ref{eq:rhows},
\ref{eq:Xws}) we obtain the general relation 
\begin{equation}
j_{w,s}\!\left(Y\right)=\left[\mathbf{p}\cdot\hat{\mathbf{n}}\,Y/m\right]_{s}\rho_{w,s}\:.\label{eq:gNEWT}
\end{equation}
Pressure due to outcoming (incoming) particles is the normal momentum
flux per unit time and area. Thus, choosing $Y=\mathbf{p}\cdot\hat{\mathbf{n}}$
to replace in Eq. (\ref{eq:gNEWT}), we obtain the pressure $P_{w,s}=j_{w,s}\!\left(\mathbf{p}\cdot\hat{\mathbf{n}}\right)=\rho_{w,s}kT_{w,s}^{(n)}$,
i.e. the expression in Eq. (\ref{eq:NEWTs}). This completes our derivation
of the Wall theorem for the pressure in out-of-equilibrium, stationary
conditions. 

Before proceeding to the validation of the NEWT with MD simulations,
we would like to point out a further generalization based on the given
derivation. On one hand, Eqs. (\ref{eq:rhows}, \ref{eq:Xws}, \ref{eq:Pws})
and (\ref{eq:NEWT}) correspond to mean values over the boundary $\partial\mathcal{A}$.
However, the identity should be valid not only for the integrals but
also for the integrands in Eqs. (\ref{eq:Xws}) and (\ref{eq:jY}),
once we replace $X\rightarrow\mathbf{p}\cdot\hat{\mathbf{n}}\,Y/m$.
Therefore, the derivations of Eqs. (\ref{eq:rhows}, \ref{eq:Xws},
\ref{eq:Pws}) and (\ref{eq:NEWT}) can be also performed for a small
patch in $\partial\mathcal{A}$ around a point $\mathbf{r}$ of the
surface, instead of using the complete boundary. This is the case,
at least, for any smooth region of $\partial\mathcal{A}$, where the
direction normal to the surface is well-defined. Under this condition,
following a similar approach to that used above, and choosing $Y=\mathbf{p}\cdot\hat{\mathbf{n}}$,
a local version of Eqs. (\ref{eq:rhows}, \ref{eq:Xws}, \ref{eq:Pws})
and (\ref{eq:NEWT}) is found. This \emph{local} NEWT for the pressure
is given by
\begin{equation}
P_{w}\left(\mathbf{r}\right)=\rho_{w}\left(\mathbf{r}\right)kT_{w}^{(n)}\left(\mathbf{r}\right)\:.\label{eq:NEWTloc}
\end{equation}
This very insteresting expression should be the subject of additional
work and it is beyond the scope of this paper.

\begin{figure}[H]
\centering{}\includegraphics[clip,width=0.75\columnwidth]{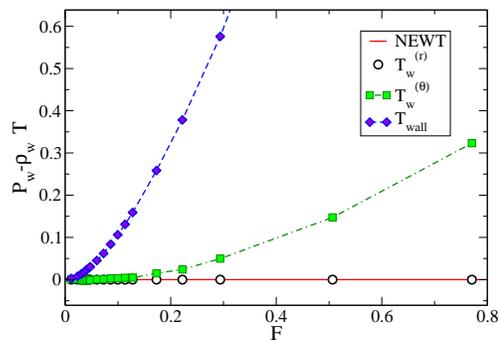}\caption{Comparison of the analytical expression of the NEWT (Eq. \ref{eq:NEWT})
and MD results, as a function of external driving $F$. Higher values
of $F$ mean more extreme non-equilibrium conditions. The red line
indicates the analytical result of Eq. (\ref{eq:NEWT}) in the form
$P_{w}-\rho_{w}T_{w}^{(r)}\equiv0$. Simulation results, in open circles,
show an excellent agreement for the complete range of external driving.
Blue diamonds correspond to the use of the wall temperature, i.e.
the equilibrium version of the Wall theorem and the curve with green
squares is obtained by using the temperaure calculated in the angular
direction.\label{fig:simproof}}
\end{figure}
Out-of-equilibrium event-driven molecular dynamics simulations were
performed to cross-check the validity of the NEWT in Eq. (\ref{eq:NEWT}).
We study a fluid flowing through a narrow cylindrical channel with
a thermostated wall. We chose event-driven molecular dynamics simulations
because they provide a precise definition and straight implementation
of hard walls, which allow for a clean comparison with the analytic
results.\citep{Klement_2019} The chosen mechanism to fix the wall
temperature is the well-known thermal-wall thermostat, described in
detail elsewhere.\citep{Tehver_1998,Taniguchi_2008} The particles
of the fluid interact through a square well potential with parameters
$\sigma=1$, $\varepsilon=1$ and $\lambda$=0.5$\sigma$. In the
axial direction, an external constant body force $F$, is applied
on each particle to induce a liquid flow. We provide more details
of the simulations in SM. We analyze the system behavior as a function
of the external force, by increasing progressivelly the flow rate
and therefore, the local viscous heating in the fluid. We point out
that the particles are thermostated only at the wall, allowing for
local heating of the bulk fluid when it is forced to flow. The mean
number density of the liquid is set to $0.6\sigma^{-3}$, and the
wall temperature to $T_{wall}=1.3\varepsilon/k$ for all the cases.
As we use reduced units for the presentation of simulation results,
they will be omitted in the following paraghraps and figures (see
SM for more details), where we also fix the Boltzmann constant to
$k=1$. The fluid temperature was measured by averaging the kinetic
energy of the particles. For simplicity, this temperature was calculated
with the kinetic energy in $\theta$ direction, in which the fluid
has no streaming velocity.

\begin{figure*}
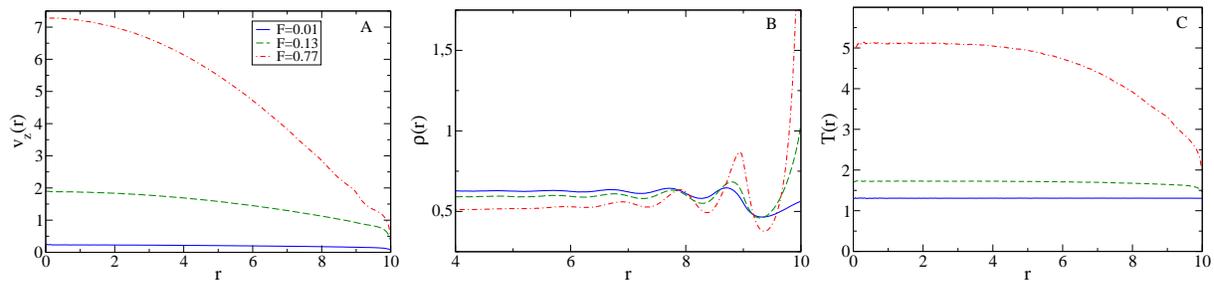

\centering{}%
\begin{tabular}{ccc}
\includegraphics[clip,width=0.22\textheight]{velrho06_pru} & \includegraphics[clip,width=0.225\textheight]{densrho06_pru} & \includegraphics[clip,width=0.22\textheight]{temprho06_pru}\tabularnewline
\end{tabular}\caption{Profiles for $v\left(r\right)$, $\rho\left(r\right)$ and $T\left(r\right)$
in the radial direction of the cylindrical nanochannel. Three volume
force values are considered, spanning the condition close-to-equilibrium
(blue curves), moderate flow (green curves) and strong out-of-equilibrium
conditions (red curves). The liquid properties change significantly
for the three cases. The velocity profile (Panel A) is close to a
parabolic Poiseuille flow, but the slip velocity on the wall increases
upon increasing flow and, at very high driving (red curves), a structure
appears, following the strong layering in the density. The liquid
density (Panel B) increases strongly close to the wall when driving
is increased and the viscous heating is more pronounced (see Panel
C).\label{fig:profiles}}
\end{figure*}
We present the comparison of the analytical NEWT expression (Eq. \ref{eq:NEWT})
with simulation results in Fig. \ref{fig:simproof}. We calculated
for the system in steady state, the mean values of pressure, density
and kinetic temperatures of the particles on the wall for different
external forces $F$. In Fig. \ref{fig:simproof} we plot the pressure
on the wall $P_{w}$ minus number density on the wall $\rho_{w}$
times characteristic temperature, i.e. an alternative expression for
the NEWT in Eq. \ref{eq:NEWT}. There, the temperature was measured
using three different prescriptions: the temperature of the wall as
fixed in the simulations, and the temperature on the wall following
two different directions of the kinetic energy in the cylindrical
nanochannel, normal to the surface $\hat{\mathbf{r}}$ and angular
$\hat{\boldsymbol{\theta}}$. The zero abcisa red line corresponds
to the analytical prediction of the NEWT from Eq. (\ref{eq:NEWT})
with normal temperature $T_{w}^{(r)}$. In this way, the proximity
to zero value indicates the level of agreement between theory and
simulation. The values of $T_{w}^{\left(r\right)}$, as calculated
from the simulations (open circles), verify the validity of the NEWT
along the entire studied range of driving forces. The simulations
provide an excelent agreement with the expression obtained for the
NEWT in Eq. (\ref{eq:NEWT}). If we use the wall temperature $T_{{\rm wall}}$
instead, it deviates from the exact behavior even for small forces.
Temperature $T_{w}^{\left(\theta\right)}$ is used as a measure of
the total temperature of the fluid at the wall. Results show that,
as the force is increased beyond $F\approx0.2$, the difference between
$P_{w}$ and $\rho_{w}T_{w}^{\left(\theta\right)}$ becomes evident,
indicating the end of the near local-equilibrium regime for the studied
system.

In SM we show MD results for the behavior of $P_{w}$ vs. $\rho_{w}$
at constant wall temperature for steady-state flow. There, the NEWT
is verified and the quality of different approximations for the temperature
at the wall are tested. In Fig. \ref{fig:profiles} the demeanor of
the confined fluid for different external forces $F$ is shown to
illustrate the general behavior of the fluid under increasingly higher
driving. From left to right panels in Fig. \ref{fig:profiles}, velocity,
density and temperature profiles are shown. As we expect, for larger
forces, higher fluxes and more pronounced velocity gradients are observed
(Panel A). At higher fluxes a significant change in the density profile
near the wall is observed (see Panel B, in Fig. \ref{fig:profiles}),
attributed to larger differences in temperatures between the central
region of the channel and the wall, due a higher viscous heating (Panel
C). This is a physical situation observed for the limit of very high
flow rate in the simulations, but expectable in nanofluidics or thermal
microdevices such as heat exchangers or dissipators. The difference
in the temperature between fluid and wall could arise from viscous
heating of the flowing liquid, as in our simulations, or because the
liquid is in contact with sources at two different temperatures. Independently
of the precise physical origin of the difference in temperatures the
NEWT is valid and relevant to shed light on the behavior of thermodynamic
quantities at the fluid-wall interface.

We point out that even when we tested the NEWT with flow of a SW model
fluid confined in a cylindrical channel, the derivation of the NEWT
is general. It applies to any fluid confined by hard walls of any
geometry and under non-equilibrium stationary states. The chosen system
allows for the accurate testing of the NEWT in relevant and complex
set of conditions, such as a curved confining wall, liquid flow and
temperature gradients arising from viscous heating. The NEWT opens
the possibility of analysing accurately the difference between equilibrium
and stationaty out-of-equilibrium states of highly confined inhomogeneous
systems.

The Wall theorem, applied up to now to fluids in equilibrium, was
extended to out-of-equilibrium conditions, and including nanoconfined
systems. This extension is demonstrated theoretically and validated
numerically using event-driven molecular dynamics under strong non-equilibrium
situations. We test our analytic results for a fluid flowing through
a small cylindrical channel with very-high temperature and velocity
gradients. We expect that this result will be relevant for theory
and experiments on fluids at the nanoscale. The NEWT should be of
particular interest in nanofluidics, in which many measurements of
the system are done at the surface of the confining media. We emphasize,
however that the result remains valid for any stationary non-equilibrium
state such as microflows and even macroscopic systems, which may involve
compressible non-newtonian fluids and could present temperature gradients,
velocity gradients and fluid flow.


\end{document}